\documentclass[12pt]{article}
\input epsf
\newcommand{\bbeta}{{\mbox{\boldmath $\beta$}}}

\newcommand{\bphi}{{\mbox{\boldmath $\phi$}}}

\newcommand{\be}{\begin{equation}}
\newcommand{\ee}{\end{equation}}
\newcommand{\bea}{\begin{eqnarray}}
\newcommand{\eea}{\end{eqnarray}}
\newcommand{\ba}{\begin{array}}
\newcommand{\p}[1]{(\ref{#1})}
\newcommand{\ea}{\end{array}}

\def\bbox{{\,\lower0.9pt\vbox{\hrule \hbox{\vrule height 0.2 cm
\hskip 0.2 cm \vrule height 0.2 cm}\hrule}\,}}
\newcommand{\dsl}{\pa \kern-0.5em /}

\newcommand{\ep}{\epsilon}

\newcommand{\nn}{\nonumber \\}

\def\Tr{{\rm Tr\,}}

\def\CM{{\cal M}}

\def\ds{\raise.15ex\hbox{/}\kern-.57em\partial}
\def\Ds{\,\raise.15ex\hbox{/}\mkern-13.5mu D}
\begin{document}

\makeatletter
\renewcommand{\theequation}{\thesection.\arabic{equation}}
\@addtoreset{equation}{section}
\makeatother

\baselineskip 18pt


\begin{titlepage}
\vfill
\begin{flushright}
QMW-PH-00-06\\
KIAS-P00048\\
hep-th/0008031\\
\end{flushright}

\vfill

\begin{center}
\baselineskip=16pt
{\Large\bf General Low Energy Dynamics \\
of Supersymmetric Monopoles} \\
\vskip 10.mm
{ ~Jerome P. Gauntlett$^{*,1}$,
~Chanju Kim$^{\dagger,2}$, Kimyeong Lee$^{\dagger,3}$ and 
Piljin Yi$^{\dagger,4}$ } \\
\vskip 0.5cm
{\small\it
$^*$
  Department of Physics, Queen Mary and Westfield College\\
  Mile End Rd, London E1 4NS, UK
}\\
\vspace{6pt}
{\small\it
 $^\dagger$
School of Physics, Korea Institute for Advanced Study\\
207-43, Cheongryangri-Dong, Dongdaemun-Gu, Seoul 130-012, Korea
}
\end{center}
\vfill
\par
\begin{center}
{\bf ABSTRACT}
\end{center}
\begin{quote}
We derive general low energy dynamics of monopoles and dyons in
$N=2$ and $N=4$ supersymmetric Yang-Mills theories by utilising a
collective coordinate expansion. The resulting new kind of supersymmetric
quantum mechanics incorporates the 
effects of multiple Higgs fields, both in the $N=2$ vector multiplet 
and hypermultiplets, having non-vanishing expectation values. 

\vfill
\vskip 5mm
\hrule width 5.cm
\vskip 5mm
{\small
\noindent $^1$ E-mail: j.p.gauntlett@qmw.ac.uk \\
\noindent $^2$ E-mail: cjkim@ns.kias.re.kr\\
\noindent $^3$ E-mail: klee@ns.kias.re.kr\\
\noindent $^4$ E-mail: piljin@ns.kias.re.kr\\
}
\end{quote}
\end{titlepage}
\setcounter{equation}{0}

\section{Introduction}

Super-Yang-Mills theories with extended supersymmetry
have a rich spectrum of BPS monopole and dyon states. 
At weak coupling one can use semi-classical techniques 
to study their properties and one finds that the low-energy
dynamics is governed by some kind of a supersymmetric quantum mechanics 
based on the moduli space of classical BPS monopole solutions.

Early work analyzed points in the classical 
moduli space of vacua of the field theory
where only a single adjoint Higgs field is non-vanishing \cite{gaunt,blum,
sen,sethi,cederwall,gauntharv,gauntlowe,lwy,gibbons}. 
In this case the electric and magnetic charge vectors 
of the BPS dyons are proportional to each other and they preserve
1/2 of the supersymmetry. More recently it has been realised that
when a second adjoint Higgs field is non-vanishing there is an
interesting spectrum of BPS states with electric and magnetic charge
vectors that are not parallel 
\cite{bergman,hashimoto,bergmankol,kawano,hhs,leeyi,bak}. 
In theories with $N=4$ supersymmetry such BPS
states preserve 1/4 of the supersymmetry, while in the theories with
$N=2$ supersymmetry they still preserve 1/2. 

In this more general situation it is becoming clear that the 
supersymmetric quantum mechanics that governs the low-energy dynamics 
includes potential terms. This has been studied in the 
$N=4$ theories in \cite{tong,blly,blyone,baklee,blytwo} 
and in \cite{gkpy} for pure $N=2$  super-Yang-Mills theories. 
The status of the derivation of these 
potential terms rests on two types of arguments. Firstly, 
rather direct arguments for the existence of a bosonic potential
\cite{blly,blyone,baklee,blytwo} 
and secondly, indirect arguments for the fermions based
on supersymmetry considerations \cite{blly,blyone,baklee,blytwo,gkpy}.

Here we will improve upon the indirect arguments 
by showing that the supersymmetric quantum mechanics can in 
fact be derived using a more direct 
collective coordinate approach generalizing that of 
\cite{gaunt,blum,cederwall}. In
addition to verifying the result of \cite{gkpy} for pure $N=2$ super-Yang-Mills
theory this approach allows us to generalize to $N=2$ theories with 
hypermultiplets when the two adjoint Higgs fields in the
$N=2$ vector multiplet are non-vanishing. 
The resulting supersymmetric quantum mechanics in this case 
both generalizes that of \cite{sethi,cederwall,gauntharv}, which only
considered a single Higgs field, and that of 
\cite{gkpy} which didn't include hypermultiplets. As the $N=4$ theory
is an $N=2$ theory with a single massless adjoint hypermultiplet, we also
recover the supersymmetric quantum mechanics presented in \cite{blly}
as a special case.

We will also consider how the supersymmetric quantum
mechanics is modified when the scalars in the hypermultiplets acquire
expectation values while maintaining a non-trivial Coulomb branch. 
In doing so we derive the general supersymmetric quantum mechanics 
for $N=4$ SYM theory presented in \cite{blytwo} when all six Higgs fields
have non-vanishing expectation values, as well as making contact 
with the models of \cite{lyexact}.

The supersymmetric quantum mechanics with potential terms that
was presented in \cite{gkpy} and generalized here are new. 
We will show that they can be obtained by a non-trivial 
``Scherk-Schwarz'' dimensional reduction 
of two-dimensional $(4,0)$ supersymmetric sigma models.

The plan of the rest of this paper is as follows: in section 2 we 
discuss pure $N=2$ Super-Yang-Mills theory. We briefly recall that
the general BPS equations consist of the usual BPS equations for
a single Higgs field plus a secondary BPS equation. 
We next review some aspects of the geometry of the moduli space of solutions 
to the BPS equation for a single Higgs field that we use later. 
This section concludes by carrying out the collective coordinate expansion
leading to the supersymmetric quantum mechanics of \cite{gkpy} 
that describes the low-energy monopole dynamics when the two 
adjoint Higgs fields have non-vanishing expectation values.

Section 3 generalizes the discussion to include matter fermions from
hypermultiplets. The zero modes of the matter fermions gives rise to
an Index bundle on the monopole moduli space. The effect of the second
Higgs field is to introduce extra terms in the supersymmetric 
quantum mechanics constructed from a two-form on this bundle.

Section 4 generalizes to cases when scalars from the 
hypermultiplets also acquire expectation values in addition to the
two adjoint Higgs fields in the $N=2$ vector multiplet. The analysis
covers the case that the hypermultiplets are in real representations
of the gauge group. Since the $N=4$
model is an $N=2$ model with a single massless adjoint hypermultiplet,
this analysis includes a derivation of the supersymmetric quantum
mechanics presented in \cite{blytwo}. 

For the convenience of the reader Section 5 summarizes various 
aspects of the general dynamics and discusses the quantization. 
Section 6 briefly concludes.
Finally, appendix A contains some technical calculations used in
the text, while appendix B shows how the supersymmetric 
quantum mechanics that we derive can be obtained via non-trivial,
``Scherk-Schwarz'', dimensional reduction.

\section{Dynamics of Monopoles in Pure $N=2$ SYM}

\subsection{BPS Equations}
The pure $N=2$ super-Yang-Mills Lagrangian is given by
\bea\label{susylag}
L&=&-\Tr\Biggl\{-{1\over 4}F_{MN}F^{MN}+{1\over 2}D_M\Phi^ID^M\Phi^I 
-{1\over 2}[\Phi^1,\Phi^2]^2 \nonumber \\ && 
\hskip 1cm - i\bar\chi\gamma^M D_M\chi + i\bar \chi[\Phi^1,\chi]
-\bar\chi\gamma_5[\Phi^2,\chi]\Biggr\} ,
\eea
where $\Phi^I$, $I=1,2$ denote the two real Higgs fields,
$D_M\Phi^I=\partial_M\Phi^I+[A_M,\Phi^I]$, $\chi$ is a Dirac spinor
and all fields are in the adjoint representation of the gauge group $G$.
The anti-hermitian generators of the Lie algebra ${\cal G}$
are normalised so that $\Tr t^a t^b=-\delta^{ab}$. Our metric
has mostly minus signature and $\gamma_5=i\gamma_0\gamma_1\gamma_2\gamma_3$.
The classical vacuum satisfy $[\Phi^1,\Phi^2]=0$ and thus $\phi^I$ lie
in the Cartan subalgebra of $G$: $\Phi^I={\bphi}^I\cdot {\bf H}$.
We will only consider vacua where the symmetry is maximally 
broken to $U(1)^r$ where $r$ is the rank of $G$. 
For a given vacuum we can define electric and 
magnetic charge two-vectors via
\bea
Q^I_e &=& -\Tr\oint \hat n\cdot \vec E \,\Phi^I={\bphi}^I\cdot {\bf q}, \nn
Q^I_m &=& -\Tr\oint \hat n\cdot \vec B \,\Phi^I={\bphi}^I\cdot{\bf g},
\eea
where the integration is over the asymptotic two-sphere with outward 
normal unit vector $\hat n$, and we have introduced the
electric and magnetic charge vectors given by
\bea
{\bf q}&=&n^m_e {\bbeta^m} ,\nn
{\bf g}&=&4\pi n^m_m{\bbeta_m^*} ,
\eea
respectively, where ${\bbeta^m}$ are the simple roots and
${\bbeta_m^*}$ are the simple co-roots of $\cal G$, and
$n^m_m$ are the topological winding numbers and $n^m_e$
are, in the quantum theory, the electric quantum numbers.   

There is a classical bound on the mass given by \cite{fh,leeyi}
\be\label{bpse}
M\ge {\rm Max}\;
\left[|\vec{Q}_e|^2 + |\vec{Q}_m|^2 \pm 2 (Q^2_mQ^1_e-Q^1_mQ^2_e)\right]
^{1/2} .
\ee
It can also be written in the form ${\rm Max}\:|Z_\pm|$ where
$Z_\pm =(Q^1_e\pm Q^2_m)+ i(Q^1_m\mp Q^2_e)$. Only the charge 
$Z_-$ appears as a central charge in the
$N=2$ supersymmetry algebra and BPS states preserving 1/2
of the supersymmetry satisfy $M=|Z_-|$ \cite{wittenolive,gkpy}. 
A consequence of the bound (\ref{bpse}) is that classical BPS solitons 
can only have charges that satisfy $|Z_-| \ge |Z_+|$. 
In subsequent sections we will mostly be concerned with BPS solitons.

The mass bound (\ref{bpse}) is saturated when
\bea\label{bpsbound}
\vec E&=&\pm \vec Da ,\nn
\vec B&=&\vec Db ,
\eea
where we have defined the rotated Higgs fields via
\bea\label{aandb}
a&=&\cos\alpha\Phi^1-\sin\alpha\Phi^2 ,\nn
b&=&\sin\alpha \Phi^1+\cos\alpha\Phi^2 ,
\eea
and the angle $\alpha$ is constrained to be 
\be\label{anglevalue}
\tan\alpha = {Q_m^1 \mp Q_e^2\over Q_m^2 \pm Q_e^1}\, .     
\ee
The second equation in (\ref{bpsbound}) is the usual
BPS equation for a single Higgs field and is referred to as the ``primary BPS
equation''. If we take static fields in the gauge $A_0=\mp a$, 
Gauss' Law  becomes the ``secondary BPS equation'' for the field $a$:
\be\label{gausslaw}
D^2a+[b,[b,a]]=0 .
\ee
For a given solution of the primary BPS equation, the secondary BPS
equation is exactly the same equation that is solved by gauge functions that
generate zero modes about the original solution. For specified asymptotic
behavior of $a$ it has a unique solution. The solutions to the general
equations can thus be viewed as electrically dressed solutions to the primary
BPS equation. Finally we note that in terms of the vectors 
${\bf a},{\bf b}$, the mass bound is given by 
\bea\label{boundalt}
M&\ge& {\rm Max}\:|i(\pm{\bf a}\cdot {\bf q} +{\bf b}\cdot{\bf g} )
                   +({\bf b}\cdot {\bf q} \mp{\bf a}\cdot{\bf g})   | \nn
&=&{\rm Max}\:(\pm{\bf a}\cdot {\bf q} +{\bf b}\cdot{\bf g} )  .
\eea
where the second expression is obtained by noting that \p{anglevalue}
can be recast as the constraint 
\be\label{constraintv}
{\bf b}\cdot {\bf q}= \pm{\bf a}\cdot{\bf g}
\ee

\subsection{Zero Modes}

As we will discuss in the next subsection,
the collective coordinate expansion is constructed about
solutions of the ordinary BPS equation for a single Higgs field 
$B_i=D_i \Phi$. It will be useful to summarize some aspects of
the discussion of the geometry of the moduli spaces of solutions
as presented in \cite{harvstrom,gaunt}.
We first define a connection $W_\mu$ on $R^4$ that is
translationally invariant in the four direction via
$W_\mu=(A_i,\Phi)$. If $G_{\mu\nu}$ is the corresponding field strength 
then the BPS equations can be recast as self duality
equations for $W_\mu$,
\be
G_{\mu\nu}={1\over 2}\epsilon_{\mu\nu\rho\sigma}G_{\rho\sigma}.
\ee
Introducing the covariant derivative
on $R^4$,
$D_\mu=\partial_\mu+[W_\mu,\ ]$,
we note that an infinitesimal gauge transformations on 
$(A_i,\Phi)$ can be recast in the form
$\delta W_\mu(x)=D_\mu\Lambda$ if the gauge parameter 
$\Lambda(x)$ is restricted to be independent of $x^4$.

Denote the moduli space of solutions to the BPS
equations within a given topological class $k$ by ${\cal M}_k$.
A natural set of coordinates is provided by the moduli ${z^m}$ that specify
the most general gauge equivalence class of solutions $W_\mu(x,z)$.
The zero modes $\delta_m W_{\mu}$
about a given solution 
satisfy the linearized self-duality equation
\be
D_{[\mu}\delta_m W_{\nu]}=
{1\over 2}\epsilon_{\mu\nu\rho\sigma}D_{\rho}\delta_m W_{\sigma},
\ee
as well as 
\be\label{orth}
D_\mu\delta_m W_\mu=0.
\ee
They can be used to construct a natural metric on 
${\cal M}_k$ via:
\be\label{metrict}
g_{mn}=-\int d^3x \Tr(\delta_m W_\mu\delta_n W_\mu).
\ee
We see that (\ref{orth}) implies that the
zero mode is orthogonal to gauge modes.

If we let $W_\mu(x,z)$ be a family of BPS monopole configurations,
the zero modes are given by 
\be
\delta_m W_\mu=\partial_m W_\mu-D_\mu \eta_m,
\ee
where the gauge parameters $\eta_m(x,z)$ are chosen to 
satisfy (\ref{orth}). The gauge parameters $\eta_m$ 
define a natural connection on $\CM_k$ with covariant derivative
\be
s_m=\partial_m+[\eta_m,\ ],
\ee
and field strength 
\be
\phi_{mn}=[s_m,s_n].
\ee
The pair $(W_\mu(x,z),\eta_m(x,z))$ defines a natural connection on
$R^4\times\CM_k$. The components of the field strength are given by
$G_{\mu\nu}$, $\phi_{mn}$ and the mixed components are given by
\be\label{cross}
[s_m,D_\mu]=\delta_m W_\mu.
\ee
Note the following identities:
\bea
s_mG_{\mu\nu}&=&2D_{[\mu}\delta_mW_{\nu]},\nn
D_\mu\phi_{mn}&=&-2s_{[a}\delta_{b]}W_\mu,\nn
\phi_{mn}&=&2(D_\mu D_\mu)^{-1}[\delta_m W_\nu,\delta_n W_\nu].
\eea

The Christoffel connection associated with the metric 
(\ref{metrict}) can be written in the form:
\be
\Gamma_{mnk}=g_{ml}{\Gamma^l}_{nk}=-\int d^3x
\Tr(\delta_m W_\mu s_k\delta_n W_\mu).
\ee
The hyper-K\"ahler structure on $R^4$ gives rise to a hyper-K\"ahler
structure on ${\cal M}_k$. The three complex structures can
be written
\be\label{compstr}
J^{(s)n}_m=-g^{np}\int d^3x {J^{(s)}}_{\mu\nu} \Tr
(\delta_m W_\mu\delta_p W_\nu),
\ee
and we note that
\be\label{hkid}
J^{(s)n}_m\delta_n W_\mu=-J^{(s)}_{\mu\nu}\delta_m W_\nu.
\ee

We now recall some aspects of the zero modes of the adjoint fermions.
It is convenient to introduce hermitian Euclidean gamma matrices via
\be
\Gamma_i=\gamma_0\gamma_i,\qquad\Gamma_4=\gamma_0,
\ee
satisfying $\{\Gamma_\mu,\Gamma_\nu\}=2\delta_{\mu\nu}$ and 
define $\Gamma_5=\Gamma_1\Gamma_2\Gamma_3\Gamma_4$. The fermion
zero modes are time independent solutions of the Dirac equation
in the presence of a BPS monopole and thus solve:
\be
\Gamma_\mu D_\mu\chi=0.
\ee
They are necessarily anti-chiral. The monopole breaks 1/2 of the supersymmetry
and they unbroken supersymmetry can be used to pair the bosonic 
and fermionic zero modes
via 
\be\label{fizz}
\chi_m = \delta_m W_\mu\Gamma^\mu\epsilon_+,
\ee
where $\epsilon_+$ is a c-number spinor that can be chosen to satisfy
\be\label{sunday}
\epsilon_+^\dagger\epsilon_+=1, \qquad
J^{(3)}_{\mu\nu}=-i\epsilon_+^\dagger\Gamma_{\mu\nu}\epsilon_+ .
\ee
Using (\ref{hkid}) we deduce that the fermionic zero modes satisfy
\be\label{fzmid}
J_m^{(3)n}\chi_n=i\chi_m ,
\ee
and hence that two bosonic zero modes
are paired with one fermionic zero mode,
in accord with the Callias index theorem \cite{callias}.

\subsection{Bosonic Monopole Dynamics}

The semi-classical quantization of BPS
monopoles begins with a mode expansion of the fields
about a given classical solution. For each
zero mode one must introduce a collective co-ordinate.
By ignoring all of the non-zero modes one obtains 
a description of the low-energy dynamics. For the case of
a single Higgs field in pure $N=2$ SYM 
this was carried out in detail in \cite{gaunt}. The resulting supersymmetric
quantum mechanics is a consistent, i.e. supersymmetric, truncation
of the full field theory dynamics. Here we generalize this 
derivation to include the effects of a second Higgs field having
a non-vanishing expectation value.

Let us first consider the bosonic case. There have been a number of separate
but related arguments that conclude that
the effect of the second Higgs field, in an appropriate limit, 
is to give rise to a potential term 
that is the norm of a tri-holomorphic Killing vector
on the moduli space \cite{blly,blyone,baklee,blytwo}. 
Let us paraphrase the arguments here in
a way that is most useful to include fermions.

We begin by emphasising that we derive the low-energy dynamics 
of monopoles; dyons then emerge as particular excited states
of the monopole dynamics. We thus begin with a given magnetic
charge vector ${\bf g}$ and fixed Higgs expectation values $\Phi^I$.
Setting ${\bf q}=0$ then fixes the angle $\alpha$ 
\p{anglevalue} and hence specifies the fields $a,b$ defined in \p{aandb}. 
It is important to notice that this means the expectation value
${\bf a}$ is orthogonal to the magnetic charge,
\be\label{adotg}
{\bf a\cdot g}=0.
\ee
The collective coordinate expansion then begins with a static purely magnetic 
solution to the primary BPS equation $B_i=D_ib$. 
The dynamical effect of the second Higgs field is treated as 
a perturbation of this solution. 
The collective coordinate expansion can be considered to be an 
expansion in the number of time derivatives $n=n_\partial$.
The equations of motion of the low-energy effective action  
will be  of order $n=2$ so we must ensure that a collective
coordinate ansatz solves 
the equations of motion of the field theory to order 
$n=0$ and $n=1$. To incorporate the affects of the second Higgs
field we will also assume that $a$ is of order $n=1$.
We next write the Lagrangian in terms of $b,a$ rather than $\Phi^1,\Phi^2$,
respectively, to obtain
\be
L=-{1\over 2}
\Tr\Biggl\{-{1\over 2}F_{MN}F^{MN}+D_M aD^M a + D_M bD^M b 
+[a,b]^2 \nonumber\Biggr\} .
\ee
To order $n=0$ the equations of motion are all solved for a
time dependent solution to the primary BPS equation $W_\mu(x,z(t))$,
with $W_4=b$. 
At order $n=1$ we need to solve the $A_0$ equation of motion,
Gauss's Law, and the $a$ equation of motion. The former is solved,
as usual, by setting $A_0=\dot z^m\eta_m$ and noting that the terms
involving $a$ are higher order. The order $n=1$ equation of motion for
$a$ is simply the secondary BPS equation $D_\mu D_\mu a=0$, 
since $D_0 D_0a$ is higher order. This
equation has a unique solution for specified asymptotic behavior 
(expectation value) of $a$. Since this is precisely the equation satisfied
by the gauge parameter specifying the gauge-zero mode, $D_\mu a$ must be a
linear combination of gauge zero modes. More precisely we have
\be\label{horse}
D_\mu a=-G^m \delta_m W_\mu ,
\ee
where $G^m$ is a linear combination of the $r$
tri-holomorphic Killing vector fields ${\bf K}$ on $\CM_k$ corresponding
to the $U(1)^r$ gauge transformations\footnote{Note that the sign
appearing in \p{horse} is related to a choice of 
conventions for the signs of the Killing vectors ${\bf K}$.}:
\be
G={\bf a}\cdot {\bf K}.
\ee
Having solved the equations of motion to order $n=0,1$ we can substitute the
ansatz into the field theory Lagrangian. After integrating over space
we get
\be\label{bosqm}
S={1\over 2}\int dt[\dot z^m \dot z^n g_{mn} - G^m G^n g_{mn}]
-{\bf b}\cdot {\bf g}.
\ee

Note that the corresponding energy admits a Bogomol'nyi bound,
$E\ge |\dot z^m G_m|+{\bf b}\cdot {\bf g}$, that is saturated 
when $\dot z^m=\mp G^m$. States saturating this bound then have energy
given by $E=G^m G^n g_{mn}+{\bf b}\cdot {\bf g}$. Using our ansatz we
next note that the electric field can be expressed via $E_i=\dot z^m \delta_m
W_i$. For configurations with  $\dot z^m=\mp G^m$ we have $E_i=\pm D_i a$.
Using the argument in \cite{tong} we can then show that the energy of
theses states can be recast in the form 
$E=\pm {\bf a}\cdot{\bf q}+{\bf b}\cdot {\bf g}$. 

To relate this to the mass formula \p{boundalt} it is
helpful to first recall that the monopole moduli space splits into
the product, modulo a discrete
identification, of a centre of masss piece with a piece describing
the relative motion of fundamental monopoles. Since the electric
charge arising from the center of mass part is
necessarily parallel to ${\bf g}$, we see that the electric 
excitation energy $\pm {\bf a}\cdot{\bf q}$ only captures the 
excitation energy due to relative electric charges. On the other hand
centre of mass sector contribution to the electric energy can be written as
$({\bf b}\cdot{\bf q})^2/2{\bf b}\cdot{\bf g}$.
Thus, in the moduli space approximation that began with 
${\bf a}\cdot{\bf g}=0$\footnote{Note that ${\bf a}\cdot{\bf g}=0$ also implies
that $G^2=0$ for the centre of mass part of the Lagrangian.}
the electric energy of a BPS dyon splits cleanly into
two pieces; $\pm{\bf a}\cdot{\bf q}$ arising from the
electric energy  of the relative sector and 
$({\bf b}\cdot{\bf q})^2/2{\bf b}\cdot{\bf g}$
from the center of mass. We see that this is consistent
with the expansion of the first line of \p{boundalt}: 
\be
M\simeq {\bf b}\cdot{\bf g} \pm {\bf a}\cdot{\bf q} 
+\frac{({\bf b}\cdot{\bf q})^2}{2{\bf b}\cdot{\bf g}},
\ee

\subsection{Supersymmetric Monopole Dynamics}

Let us now turn to a derivation of effective action when we include
the fermions in the pure $N=2$ super-Yang-Mills theory.
It is again convenient to rewrite the 
pure $N=2$ super Yang-Mills action in terms of $a,b$. Noting that
$(a,b)$ is a rotation of $(-\Phi^2,\Phi^1)$ we obtain
\bea\label{susylagab}
L&=&-\Tr\Biggl\{-{1\over 4}F_{MN}F^{MN}+{1\over 2}D_M aD^M a + 
{1\over 2}D_M bD^M b 
-{1\over 2}[a,b]^2 \nonumber \\ && 
\hskip 1cm - i\bar\chi\gamma^M D_M\chi +i\bar \chi[b,\chi]
+\bar\chi\gamma_5[a,\chi]\Biggr\} ,
\eea
with it understood that $\chi$ has now been rotated by the angle 
$(\alpha-\pi/2)/2$.
The collective coordinate expansion can now be
considered to be an expansion in  $n=n_\partial+{1\over 2}n_f$.
where $n_f$ as the number of fermions. 
A low-energy ansatz for the fields
should solve the equations of motion to order
order $n=0, {1\over 2},1$. By combining the ansatz for the case of a single
Higgs field in \cite{gaunt} with the above ansatz for the bosonic case
we are led to
\bea\label{ansatztwo}
W_\mu&=&W_\mu(x,z(t)),\nn
\chi&=&\delta_mW_\mu\Gamma^\mu\epsilon_+\tilde\lambda^m(t),\nn
A_0&=&\dot z^m\eta_m-i\phi_{mn}\tilde\lambda^{\dagger m}\tilde\lambda^n,\nn
a&=&\bar a +i\phi_{mn}\tilde\lambda^{\dagger m}\tilde\lambda^n,
\eea
with 
\be\label{cat}
D_\mu \bar a = -G^m\delta_m W_\mu .
\ee
Because of (\ref{fzmid}) 
the complex fermionic Grassmann odd collective coordinates
$\tilde\lambda^m$ are not independent and satisfy
\be\label{fccid}
-i\tilde\lambda^m J_m^{(3)n}=\tilde\lambda^n .
\ee
Real independent $\lambda^m$ can be defined via
\be \label{reallambda}
\lambda^m=\sqrt{2}\left(\tilde\lambda^m+(\tilde\lambda^m)^\dagger\right) .
\ee

If we ignore the shift in $a$ by $\bar a$, we have the ansatz 
for the case of a single Higgs field analysed in \cite{gaunt}.
Hence after substituting into the action (\ref{susylagab})
the $\bar a$ independent terms lead to the supersymmetric quantum
mechanics:
\be\label{actionone}
S={1\over 2}\int dt[\dot x^m \dot x^n g_{mn}+
ig_{mn} \lambda^m D_t \lambda^n ] 
-{\bf b}\cdot {\bf g},
\ee
where
\be
D_t\lambda^m=\dot \lambda^m
+\Gamma^m_{nk}\dot z^n\lambda^k .
\ee
Since the $\bar a$ dependent terms arising 
from $(D_0a)^2$ in the action are again higher order than we are
considering, we just need to focus on $D_\mu a D_\mu a$ and 
the Yukawa terms $\bar\chi \gamma_5[a,\chi]$. The $\bar a$
dependent terms in the former are
\be
{1\over 2} \Tr D_\mu\bar a D_\mu\bar a
+i\tilde\lambda^{\dagger m}\tilde\lambda^n \Tr D_\mu(\phi_{mn} 
D_\mu \bar a) .
\ee
When we integrate over the spatial coordinates the second term vanishes
and we are left with bosonic potential
\be\label{pot}
-{1\over 2} G^m G^n g_{mn} .
\ee
as in the bosonic case.
The $\bar a$ terms arising from the Yukawa term give rise to
\be
i\int d^3 x \Tr \chi^\dagger[\bar a,\chi] ,
\ee
which can be rewritten as
\bea
&&-2i \tilde\lambda^{\dagger m}\tilde\lambda^n \int 
d^3 x\Tr \delta_m W_\mu[\bar a,
\delta_n W_\mu]\nn
&=&-2i\tilde\lambda^{\dagger m}\tilde\lambda^n \int d^3 x\Tr 
\delta_m W_\mu (D_\mu s_n \bar a -s_n D_\mu\bar a) ,
\eea
where we have used (\ref{sunday}) and (\ref{cross}), respectively.
Using the fact that
\be\label{monday}
s_m\bar a = G^n\phi_{mn},
\ee
which can be proved by acting on both sides with $D^2$ and using the
fact that $D^2$ has no zero modes, we note that the first term is
a boundary term which vanishes. 
The remaining term can then be recast
in the form
\be\label{fbilin}
2i\tilde\lambda^{\dagger m}\tilde\lambda^n G_{m;n}={i\over 2}\lambda^m
\lambda^n G_{m;n} .
\ee
In summary, the effect of the second field is thus
to add the potential term (\ref{pot}) and the fermion bilinear (\ref{fbilin})
to the action (\ref{actionone}) to obtain 
\be\label{actiontoo}
S={1\over 2}\int dt[\dot x^m \dot x^n g_{mn}+
ig_{mn} \lambda^m D_t \lambda^n - G^m G^n g_{mn}
- iD_m G_n  \lambda^m \lambda^n] 
-{\bf b}\cdot {\bf g} .
\ee
We have thus derived the supersymmetric quantum mechanics that
was first presented, based on supersymmetry considerations,
in \cite{gkpy}.

\section{Inclusion of Matter Fermions}

We now consider the low-energy dynamics of monopoles in
$N=2$ Yang-Mills theories with hypermultiplets. This was first
studied in \cite{sethi,cederwall,gauntharv} in the special case
that only a single adjoint Higgs field has a non-trivial expectation value.
The main new feature is that the matter fermions give rise to 
extra fermionic zero modes that provide a natural Index bundle over
the moduli space of monopoles. The resulting supersymmetric
quantum mechanics is coupled to this bundle. Here we will show
that when the second adjoint Higgs field of the $N=2$ vector 
multiplet has a non-vanishing expectation value, this supersymmetric 
quantum mechanics is modified by terms 
constructed from a natural two-form on this bundle.

The massless hypermultiplet contribution to the Lagrangian is given by
\bea\label{matterlag}
L_H&=&\frac12 D_K M^{\dagger } D^K  M 
      + i \bar\Psi\gamma^K D_K \Psi 
      - \bar\Psi(-i\Phi_1 - \gamma_5 \Phi_2)\Psi \nn
   &&+ M^{\dagger 1}\bar\chi\Psi + \bar\Psi\chi M_1
     + iM^{\dagger 2}\bar\chi^c\gamma_5\Psi 
     + i\bar\Psi\gamma_5\chi^c M_2 \nn
   &&+\frac12 M^{\dagger} (\Phi_1^2 + \Phi_2^2) M
     +\frac18 (M^{\dagger } t^\alpha \tau_s M)^2,
\eea
where $M$ is a doublet of complex scalars $(M_1,M_2)^T$,
$t^\alpha$ are anti-hermitian generators in the matter 
representation, $\tau_s$ are Pauli matrices, and 
$\chi^c$ is the charge conjugation of $\chi$ (defined precisely 
in section 4).

\subsection{Zero Modes and the Index Bundle}

Before discussing the effects of the second adjoint Higgs field,
let us briefly discuss some of the geometry of the Index bundle defined
by the fermion zero-modes. The fermion zero modes solve 
the Dirac equation in the background of a monopole configuration
\be\label{diraceq}
\Gamma_\mu D_\mu\gamma_5\Psi =0    ,
\ee
and are chiral. Let $\Psi_A(x,z)$, $A=1\dots l$ be a basis of the
fermion zero modes in monopole background specified by the moduli $z$
satisfying
\be\label{dog}
\int d^3 x \Psi^{\dagger}_{\bar A}\Psi_B\equiv<\Psi_{\bar A}|\Psi_B>=
\delta_{\bar AB}  ,
\ee
where we have defined $\Psi^\dagger_{\bar A}\equiv(\Psi_A)^\dagger$.
It will be very useful to note the completeness relationship
\be\label{completeness}
|\Psi_A>\delta^{A \bar B}<\Psi_{\bar B}|+\Pi +{1-\Gamma_5\over 2}=1 ,
\ee
where the operator $\Pi$ projects onto the chiral non-zero modes and
has the form
\be
\Pi=\gamma_5{\Ds}{1\over D_\mu D_\mu}{\Ds}\gamma_5{1+\Gamma_5\over 2} .
\ee 
A connection on the Index bundle is defined
by
\be\label{mattercon}
{A_m}_{\bar AB}=<\Psi_{\bar A}|s_m\Psi_B> .
\ee
Using the results of section 2.2 and (\ref{completeness}) 
one can show that the corresponding
field strength can be written in the form \cite{cederwall}
\be
F_{mn\bar AB}=<s_m\Psi_{\bar A}|\Pi s_n\Psi_B>-<s_n\Psi_{\bar A}|\Pi s_m\Psi_B>
+<\Psi_{\bar A}|\phi_{mn}\Psi_B> .
\ee

It is straightforward to see that the connection one-form is unitary and
hence the structure group of the Index bundle is generically $U(l)$.
The Index bundle thus admits a covariantly constant complex
structure $I^{(3)}$ with K\"ahler form taken to be 
$I^{(3)}_{A\bar B}=i\delta_{A \bar B}$ (the superscript will be convenient
in section 4).
When the representation of the matter fermions in the gauge group is 
real or pseudo-real, however, the structure group is further restricted 
\cite{mantonschroers,lyexact}. For the pseudo-real representation, the 
structure group  of the bundle reduces to $O(l)$, while,
for the real representation, the structure group reduces to a 
symplectic bundle $USp(l)$. A special case of the latter is 
the adjoint fermion zero modes that live 
in the co-tangent bundle of the moduli space which, being hyper-K\"ahler,
is indeed symplectic. In this case note that
the field strength $F$ is simply the Riemann curvature tensor.

\subsection{Collective Coordinate Expansion}

The collective coordinate expansion with two adjoint
Higgs fields and matter fermions parallels what was done 
for the case of pure $N=2$ SYM in section 2. We again first perform
a chiral rotation to write the action in terms of $a,b$ which requires
that we work with rotated fermions and matter fields.
The ansatz for the vector multiplet fields is then given by
\bea\label{ansatzthree}
W_\mu&=&W_\mu(x,z(t)),\nn
\chi&=&\delta_mW_\mu\Gamma^\mu\epsilon_+\tilde\lambda^m(t),\nn
A_0&=&\dot z^m\eta_m-i\phi_{mn}\tilde\lambda^{\dagger m}\tilde\lambda^n
+\frac{i}{D^2}(\Psi^\dagger t^\alpha\Psi t^\alpha),\nn
a&=&\bar a +i\phi_{mn}\tilde\lambda^{\dagger m}\tilde\lambda^n
+\frac{i}{D^2}(\Psi^\dagger t^\alpha\Psi t^\alpha),
\eea
while for the matter fields it is given by
\bea\label{ansatzfour}
\Psi&=&\psi^A(t)\Psi_A ,\nn
M_1&=&-{2\over D^2}(\bar\chi\Psi) ,\nn
M_2&=&-{2i\over D^2}(\bar\chi^c\gamma_5\Psi) ,
\eea
where we have introduced the Grassmann odd complex collective coordinates
$\psi^A(t)$ for the matter fermion zero modes. This ansatz solves the equations
of motion to order $n=0,1/2,1$ and generalises
that in \cite{cederwall} by simply shifting the $a$ field
by a gauge function $\bar a$ satisfying {\p{cat}}.

After substituting this ansatz into the field theory action,
the $\bar a$ independent terms give rise to the supersymmetric
quantum mechanics presented in \cite{sethi,cederwall,gauntharv}:
\bea\label{aaa}
{\cal L}={1\over 2} \left( g_{mn} \dot{z}^m \dot{ z}^n +
ig_{mn} \lambda^m D_t \lambda^n 
+i\psi^a{\cal D}_t\psi^a  + {1\over 2}F_{mn ab}\lambda^m
\lambda^n\psi^a\psi^b \right)-{\bf b\cdot g} ,
\eea
where
\be
{\cal D}_t\psi^a=\dot \psi^a
+{{A_m}^a}_b\dot z^m\psi^b ,
\ee
and we traded off complex $\psi^A$'s in favor of real $\psi^a$'s (effectively
this means we are embedding the $U(l)$ bundle in an $So(2l)$ bundle).
The $\bar a$ dependent terms give rise to the potential terms
presented in the last section
plus an additional fermion bilinear. This latter term can be
rewritten 
\be\label{holland}
-i\psi^{\bar A}\psi^B T_{\bar AB} ,
\ee
where we have defined $\psi^{\bar A}$ as the complex conjugate of 
$\psi^{A}$ and
\be\label{tee}
T_{\bar AB}=
<\Psi_{\bar A}|\bar a\Psi_B> .
\ee
As $T$ is anti-hermitian, in a real basis \p{holland} becomes
$-i \psi^a\psi^b T_{ab}/2$ with $T_{ab}=-T_{ba}$.
For hypermultiplets in general representations we cannot write $T$
in a simpler form. However, as we will discuss it is crucial for
consistency of the supersymmetric quantum mechanics that
\be\label{tues}
T_{\bar AB;m}= F_{mn \bar AB}G^n .
\ee
To prove this we begin with
\be\label{rain}
\partial_m T_{\bar AB}=<s_m\Psi_{\bar A}|\bar a\Psi_B> + 
<\Psi_{\bar A}|(s_m\bar a)\Psi_B> + <\Psi_{\bar A}|\bar a s_m\Psi_B> .
\ee
Using (\ref{completeness}) the first term in (\ref{rain}) can then be written
\bea
<s_m\Psi_{\bar A}|\Psi_C>\delta^{C\bar C}<\Psi_{\bar C}|\bar a\Psi_B> 
+ <s_m\Psi_{\bar A}|\Pi\bar a\Psi_B> .
\eea
The first term is $-A_{m\bar A C}\delta^{C\bar C} T_{\bar CB}$. 
Using the identity
\be
\Ds\gamma_5(\bar a\Psi_{A} - G^m s_m\Psi_A)=0 ,
\ee
which can be proven by acting with $G^m s_m$ on $\Ds \gamma_5 \Psi_A=0$,
we can rewrite the second term as
\be
G^n<s_m\Psi_{\bar A}|\Pi s_n\Psi_B> .
\ee
The last term in (\ref{rain}) 
can be manipulated in a similar way. The second term
can be rewritten using (\ref{monday}). 
Putting this together we deduce
that
\be
\nabla_m T_{\bar AB}=
G^n\{<s_m\Psi_{\bar A}|\Pi s_n\Psi_B>-<s_n\Psi_{\bar A}|\Pi s_m\Psi_B>+
<\Psi_{\bar A}|\phi_{mn}\Psi_B>\} .
\ee
Since the last term in braces is precisely the curvature
$F_{mn\bar A B}$ we have established (\ref{tues}).

In conclusion the supersymmetric quantum mechanics describing the low-energy
dynamics of monopoles in $N=2$ theories with matter when both adjoint
Higgs fields are non-vanishing is given by
\bea
{\cal L}&=&{1\over 2} \biggl( g_{mn} \dot{z}^m \dot{ z}^n +
ig_{mn} \lambda^m D_t \lambda^n
 - g^{mn} G_m G_n - iD_m G_n  \lambda^m \lambda^n \nn
&&+i\psi^a{\cal D}_t\psi^a + {1\over 2}F_{mn ab}\lambda^m
\lambda^n\psi^a\psi^b - i T_{ab}\psi^a\psi^b\biggr) -{\bf b\cdot g} .
\label{action}
\eea
The main new feature is the presence of the two-form $T$ on the Index bundle.
The action is invariant under the supersymmetry transformations
\bea
\delta z^m &=& -i\ep\lambda^m +i\ep_s {J^{(s)m}}_n \lambda^n ,\nn
\delta \lambda^m&=&(\dot z^m -G^m)\ep +{J^{(s)m}}_n(\dot z^n -
G^n)\ep_s
-i\ep_s \lambda^k \lambda^n {J^{(s)l}}_k \Gamma^m_{ln}\nn
\delta\psi^a&=&-{{A_m}^a}_b\delta z^m\psi^b ,
\eea
where $\ep,\ep_s$ are constant one component Grassmann odd parameters,
provided that in addition to the usual requirements that the
moduli space is hyper-K\"ahler and that the field strength $F$ is of
type (1,1) with respect to all complex structures,
the two form $T$ satisfies \p{tues}.
The action is also invariant under the following symmetry transformation
generated by the tri-holomorphic Killing vector $G$
\bea
\delta z^m &=&k G^m ,\nn
\delta\lambda^m&=&k{G^m}_{,n}\lambda^n ,\nn
\delta\psi^a&=&k {T^a}_b\psi^b-A^a_{m b}\delta z^m\psi^b .
\eea
where $k$ is a constant.
In the case of $N=4$ supersymmetry, i.e. a single hypermultiplet
in the adjoint representation, the bundle is the tangent bundle 
and $T_{ab}=G_{a;b}$. We have thus derived the action first presented
in \cite{blly}, which was obtained there via symmetry arguments.

\subsection{Massive Matter Fields}

Let us briefly consider the case that the hypermultiplets are 
massive\footnote{Early work on this issue can be found in \cite{henning}.}.
The relevant mass terms are given by\footnote{Since we work with the rotated 
fields $a,b$, we interpret $(m_R,m_I)$ to have been similarly rotated.}.
\be\label{masspert}
m_R\bar\Psi\Psi - m_I i\bar \Psi\gamma_5\Psi .
\ee
Recall that the collective coordinate expansion begins by writing
the field theory Lagrangian in terms of $a,b$. 
We can treat this term as a perturbation by taking the bare mass to be 
order $n=1$, i.e., the same order of magnitude as $a$ and hence smaller than
$b$. To leading order the Dirac equation for the matter 
fermions is then not modified from
\p{diraceq}.
Substituting our ansatz (\ref{ansatzfour}) into \p{masspert}, we find
that only $m_I$ part contributes;
\be\label{massfterm}
m_I\psi^{\bar A}\psi^B \delta_{\bar A B} .
\ee
In terms of real fermions, $\psi^a$, $a=1,\dots,2l$, we get
\be
{i\over 2}m_I\; \psi^a I^{(3)}_{ab}\,\psi^b . \label{mass}
\ee
This term is naturally incorporated in the supersymmetric quantum
mechanics (\ref{action}) by adding it to
$T_{ab}$, since the differential condition on $T$ allows a shift of $T$
by a covariantly constant piece.

When we quantise the supersymmetric quantum mechanics the term
\p{massfterm} will contribute a term $N_f m_I$ to the Hamiltonian
where $N_f$ is the hypermultiplet fermion number. Recalling the discussion
at the end of section 2.3, this will lead to the mass 
of the BPS sates of the supersymmetric quantum mechanics being given by
\be\label{msim}
M\simeq {\bf b}\cdot{\bf g} - {\bf a}\cdot{\bf q} 
+\frac{({\bf b}\cdot{\bf q})^2}{2{\bf b}\cdot{\bf g}}+ N_fm_I,
\ee
This result is in precise accord with the BPS mass formula arising 
from the general ${\cal N}=2$ central charge formula. 
The latter can be written
\be
M=|i({\bf b}\cdot{\bf g} - {\bf a}\cdot{\bf q} + N_fm_I)
+({\bf b}\cdot{\bf q}  + {\bf a}\cdot {\bf g}+ N_fm_R)|,
\ee
which reduces to \p{msim} in the moduli space approximation in which
${\bf a\cdot g}=0$ and $m_R$ is neglected compared to $b$.

\section{More Potentials from the Matter Sector}

In this section we analyse situations when one can
turn on additional scalar vevs in the hypermultiplets while leaving
the $U(1)$ gauge symmetries of the Coulomb phase intact. 
This will lead to additional potential terms in the supersymmetric
low-energy dynamics of the monopoles.
Considering the potential terms in the matter Lagrangian \p{matterlag}, 
we see that this is possible when the matter representation contains a 
zero-weight vector. Moreover it is only possible
when the hypermultiplets are massless. A trivial example 
is when the hypermultiplets are in the adjoint representation. 
Less trivial examples are, for instance, symmetric tensors 
for $SO(k)$ and anti-symmetric tensors for $Sp(k)$.

We will further assume in this section that the representation is real.
In this case the Index bundle associated with the matter fermions has a 
symplectic structure group and is equipped with three covariantly
constant complex structures, $I^{(s)}$.
A special case is when we have a single massless adjoint
hypermultiplet, whose zero modes live in cotangent bundle with complex
structures $I^{(s)}=J^{(s)}$, $s=1,2,3$. The field theory is 
then $N=4$ Yang-Mills theory, so our derivation of the low-energy
dynamics will include a derivation, {\it en-passant}, of the effective
action for $N=4$ monopoles that was first presented, based on
symmetry considerations, in \cite{blytwo}.

\subsection{Bosonic potential}

The effect on the monopole dynamics 
of allowing the scalar fields $M$ to acquire
expectation values is determined in a very similar manner to
the treatment of the second adjoint Higgs field $a$ in
sections 2 and 3. We regard the vevs of the two complex
scalars $M$'s as a perturbation of order $n=1$ and 
perform a perturbative expansion.

A low-energy ansatz that solves the equations of motion to order 
$n=1$ is obtained by shifting the ansatz \p{ansatzfour} via
\bea\label{ansatzsix}
\Psi&=&\psi^A\Psi_A ,\nn
M_1&=&\bar M_1-{2\over D^2}(\bar\chi\Psi) ,\nn
M_2&=&\bar M_2-{2i\over D^2}(\bar\chi^c\gamma_5\Psi) ,
\eea
where $\bar M_{1,2}$ are order $n=1$ and solve 
the covariant Laplace equation in the monopole 
background
\be\label{fact}
D^2 \bar M_{1,2}=0 .
\ee
The new terms that arise from this shift after substituting into
the field theory action are either linear or quadratic
in $\bar M_{1,2}$. The linear pieces generate fermionic bilinears and are
discussed in the next subsection, while the quadratic pieces correspond to 
bosonic potential terms.

It will be convenient to exchange the two complex scalars 
$M_{1,2}$, for four real $H_i$'s via
\bea
&&M_1 = H_3 + i H_0 , \nn
&&M_2 = - H_1 +i H_2 .
\eea
and similarly exchange $\bar M_{1,2}$ for four real $\bar H_i$'s.
Next note that, given (\ref{fact}), $\Ds \bar H_i\epsilon_+$ 
is a fermion zero mode and hence can be expanded in terms of
our basis:
\be\label{fermrel}
\Ds \bar H_i\epsilon_+ =-i\gamma^5\sqrt{2} K_i^A(z)\Psi_A\label{DH} .
\ee
The quantities $K_i^A(z)$ define four sections on the dual of the
Index bundle over the monopole moduli space.

After substituting the ansatz (\ref{ansatzsix}) into the 
field theory action and using (\ref{fermrel}) 
we find that the bosonic part of order $n=2$ that 
involves $H$ is given by 
\be
\frac12\int d^3x\;(D_\mu  \bar H_i)^\dagger(D_\mu  \bar H_i)
=|K_i^A|^2 =\frac{1}{2}K_{ia}K_i^a ,
\ee
where we rewrote the complex quantities $K_i^A$ in terms of 
real quantities $K_i^a$ by expanding
\be \label{realk}
K_i^A=\frac{1}{\sqrt{2}}\left( K_i^{2A-1}+iK_i^{2A}\right) .
\ee
Since $i$ runs from 0 to 3, there could be four
such bosonic potentials.

\subsection{Fermion bilinear terms}

After substituting \p{ansatzsix} into \p{matterlag} one finds that
the fermionic bilinear terms arising from the kinetic terms of the $H$'s
vanish.
The non-zero fermionic bilinear terms arise from the Yukawa 
couplings in \p{matterlag}. Since the derivation is reasonably long,
we point out here that the key results are given in 
\p{holo} and \p{lastequation}. 

For fermions in a real representation of the gauge group, it is often
convenient to introduce symplectic Majorana fermions:
$\tilde\Psi$ and $\tilde\chi$ are each a doublet of 
Dirac spinors defined by
\be
\tilde\chi = \pmatrix{\chi \cr -i\gamma_5\chi^c} ,\qquad
\tilde\Psi = \pmatrix{\Psi \cr -i\gamma_5\Psi^c} .
\ee
The Yukawa terms in \p{matterlag} can then be written compactly as
\be \label{ykw}
i\int d^3x\:\bar{\tilde\Psi} \tau_i \tilde\chi  H_i ,
\ee
where $\tau_i = (1, -i \tau_s)$ and $\bar H_i$ are real. 
The charge-conjugation of the spinor, $\chi$, is defined as 
\be
\chi^c\equiv C\bar\chi^T =C(\gamma^0)^T\chi^*
\ee
and similarly for $\Psi^c$,
where the charge-conjugation matrix $C$
satisfies,
\be
CC^*=-1, \qquad C\gamma_M^T=-\gamma_MC .
\ee
It follows that $C\Gamma_\mu^T=-\Gamma_\mu C$.

Accordingly, the zero mode ansatz for $\tilde\chi$ is given by
\be
\tilde\chi 
  = \pmatrix{\tilde\lambda^m\delta_m W_\mu \Gamma_\mu \epsilon_+ \cr
             \tilde\lambda^{\dagger m}\delta_m W_\mu \Gamma_\mu \epsilon'_+} ,
\ee
where $\epsilon'_+ \equiv C\epsilon_+^*$.
Because $\chi$ (and $W$) is in a 
real representation of the gauge group, complex conjugated zero modes can 
expressed as a linear combination of original zero modes:
\be \label{cmn}
\delta_m W_\mu \Gamma_\mu \epsilon'_+ =
{\cal C}_m^{\;\;\;k}\delta_k W_\mu \Gamma_\mu \epsilon_+ .
\ee
By a basis redefinition, the matrix ${\cal C}$ can be chosen to be 
anti-symmetric and unitary so that ${\cal C}^2=-1$. 
By taking the complex conjugate of the expression
$J_m^{(3)k}\chi_k=i\chi_m$, it follows that 
${\cal C}$ anticommutes with $J^{(3)}$;
\be 
{\cal C}J^{(3)}=-J^{(3)}{\cal C} .
\ee
This matrix $\cal C$ generates a second complex structure on the
moduli space which we will also denote by $J^{(2)}$.
Defining $J^{(1)}$=$J^{(2)}J^{(3)}$ we obtain the hyper-K\"ahler 
structure of the monopole moduli space
(which can be taken to be the same as \p{compstr} by an appropriate choice
of complex structures on $R^4$).
We can use \p{cmn} to give an alternate expression for the 
zero mode ansatz of $\tilde\chi$
where the roles of $\epsilon$ and $\epsilon'$ are exchanged,
\be
\tilde\chi 
  = \pmatrix{-\tilde\lambda^m{\cal C}_m^{\;\;\;k} 
\delta_k W_\mu \Gamma_\mu \epsilon'_+ \cr
   \tilde\lambda^{\dagger m}{\cal C}_m^{\;\;\;k} 
\delta_k W_\mu \Gamma_\mu \epsilon_+} .
\ee

Using $\delta_m W_\mu = [s_m, D_\mu]$, we thus
find two possible expressions
for $\chi H$
\bea \label{chih}
\chi \bar H_i &=& \tilde\lambda^m s_m \Ds \bar H_i \epsilon_+   +\dots \nn
         &=& -\tilde\lambda^m {\cal C}_m^{\;\;\;k} s_k 
             \Ds \bar H_i \epsilon'_+ +\dots ,
\eea
and also for $-i\gamma_5\chi^c H$
\bea \label{chich}
-i\gamma_5\chi^c \bar H_i &=& \tilde\lambda^{\dagger m} s_m 
\Ds \bar H_i \epsilon'_+ 
                      + \dots \nn
                   &=& \tilde\lambda^{\dagger m} {\cal C}_m^{\;\;\;k} s_k 
                      \Ds \bar H_i \epsilon_+ + \dots .
\eea
The ellipsis denote terms of the form $\Ds(\dots)$, which do not contribute  
any new terms in the low energy dynamics, once we use the Dirac equation for
$\Psi$ and the fact that the zero modes of $\Psi$ are chiral with respect 
to $\Gamma_5$. They  will be ignored subsequently.

The matter fermion zero mode ansatz $\tilde\Psi$ takes the form
\be\label{dhep}
\tilde\Psi 
=\pmatrix{\psi^{A}\Psi_A\cr -i\gamma_5\psi^{\bar A}\Psi_{\bar A}^c} .
\ee
We also have
\bea \label{dhem}
\Ds \bar H_i \epsilon_+ &=&-i\gamma_5\sqrt{2}K_i^A(z)\Psi_A ,\nn
\Ds \bar H_i \epsilon'_+ &=&  -\sqrt{2} K_i^{\bar A}(z)\Psi_{\bar A}^c ,
\eea
where the second equation is derived from the first 
(which is just \p{fermrel}).

When determining the contributions from the Yukawa couplings, 
one has a choice of 
writing out the zero modes $\chi_m$ as
$\delta_m W_\mu\Gamma^\mu \epsilon_+$ or equivalently as
${\cal C}_m^{\;\;\;k}\delta_k W_\mu\Gamma^\mu \epsilon'_+$.
Of course the answer should not depend on such choices, but the 
expression one gets does depend on the choices. In fact, we also could
rewrite the same expression based on a different $\epsilon$ associated
with different complex structures such as 
$\epsilon_+ + i \epsilon'_+$. This redundancy of expressions 
gives us a very important constraint on the quantities $K_i$. 
As will be shown in Appendix, it implies a holomorphicity condition
on $K_i$'s;
\be \label{holo}
(J^{(s)}\nabla)(I^{(s)}K_i)= \nabla K_i ,
\ee
for $s=1,2,3$. $s$ labels the three complex structure on the tangent
and the Index bundles. This fact will be used crucially in the derivation
of fermion bilinears.

In the following we are going to switch between the above 
two expansions, so that $\Psi$ is
always paired up with $\Ds H\epsilon_+$ while $\Psi^c$ is
always paired up with $\Ds H\epsilon'_+$. This can be achieved by using the
first line of \p{chih} and \p{chich} for the Yukawa terms involving
$\bar H_0$ and $\bar H_3$, and using the second line
for the Yukawa terms involving $\bar H_1$ and $\bar H_2$.

Consider first the Yukawa terms containing $\bar H_{0}$. 
One term is
\be
i \sqrt{2}\int d^3x\:(\psi^{\bar A})\Psi_{\bar A}^\dagger 
\;(\tilde\lambda^m s_m) 
\;\Gamma_5  K_0^B(z)\Psi_B +\dots ,
\ee
where the ellipses denote the second term arising from the charge conjugate. 
The two terms can then be written
\be 
-i\sqrt{2}(\tilde\lambda^m \nabla_m)(K_{0\bar A}\psi^{\bar A}) 
-i\sqrt{2}((\tilde\lambda^m)^\dagger \nabla_m)(K_{0A}\psi^{A}) 
\ee
For the Yukawa terms containing $\bar H_3$, one gets the same
two terms multiplied by $-i$ and $i$, respectively, to give
\be 
-\sqrt{2}(\tilde\lambda^m \nabla_m)(K_{3\bar A}\psi^{\bar A}) 
+\sqrt{2}((\tilde\lambda^m)^\dagger \nabla_m)(K_{3A}\psi^{A}) .
\ee

These expressions can be recast in a more useful form using
the fact that 
\bea 
((\tilde\lambda^m)^\dagger \nabla_m)(K_{0\bar A}\psi^{\bar A}) =0=
(\tilde\lambda^m \nabla_m)(K_{0A}\psi^{A}) , && \nn
((\tilde\lambda^m)^\dagger \nabla_m)(K_{3\bar A}\psi^{\bar A}) =0=
(\tilde\lambda^m \nabla_m)(K_{3A}\psi^{A}) .&& 
\eea
This can be derived using \p{holo} and the fact that
the relation among $\tilde \lambda$'s \p{fccid} implies
\bea
\tilde\lambda^m \nabla_m &=&
\frac{1}{2}\tilde\lambda^m\left(\delta_m^{\;\;\;k}-iJ_m^{(3)k}
\right)\nabla_k, \nn
(\tilde\lambda^m)^\dagger \nabla_m &=&
\frac{1}{2}(\tilde\lambda^m)^\dagger\left(\delta_m^{\;\;\;k}+iJ_m^{(3)k}
\right)\nabla_k .
\eea
The operators $(1\mp i J^{(3)})\nabla$ are holomorphic and anti-holomorphic
covariant derivatives, so $\tilde\lambda^m \nabla_m $ is composed of 
holomorphic derivatives only, while $(\tilde\lambda^m)^\dagger 
\nabla_m$ is composed of anti-holomorphic derivatives only. 
Using this the terms arising from the $\bar H_0$ Yukawa term can be written
\bea
-i\sqrt{2}(\tilde\lambda^m +(\tilde\lambda^m)^\dagger) \nabla_m
(K_{0A}\psi^{A}+ K_{0\bar A}\psi^{\bar A}) ,
\eea
while those from the $\bar H_3$ Yukawa term become
\bea
\sqrt{2}(-\tilde\lambda^m +(\tilde\lambda^m)^\dagger)\nabla_m
(K_{3A}\psi^{A}+K_{3\bar A}\psi^{\bar A}) .
\eea
We next use \p{reallambda} to write the expressions in terms of
the real and independent 
$\lambda$'s to get
\bea
-i(\lambda^m \nabla_m)(K_{0A}\psi^{A}+
K_{0\bar A}\psi^{A})
+i(\lambda^m J_m^{(3)k}\nabla_k)(K_{3A}\psi^{A}+
K_{3\bar A}\psi^{A}) .
\eea
As the final step, we trade off complex $K$'s and $\psi$'s 
in favor of real ones and find
\bea
-i(\lambda^m \nabla_m)K_{0a}\psi^a
+i(\lambda^m J_m^{(3)k}\nabla_m) K_{3a}\psi^{a} ,
\eea
as the fermion bilinears arising from $\bar H_0$ and $\bar H_3$ Yukawa terms.

The action of $-i\tau_{1,2}$ exchanges $\chi$ and $-i\gamma_5\chi^c$, 
so the $\bar H_{1,2}$ Yukawa terms are a bit different. Expanding $\chi$
in terms of ${\cal C}_m^{\;\;\;k}\delta_k W_\mu\Gamma^\mu \epsilon'_+$ 
instead, we find
\be
-\sqrt{2}((\tilde\lambda^m)^\dagger {\cal C}_m^{\;\;\;k}\nabla_k)
(K_{1\bar A} \psi^{\bar A}) 
+\sqrt{2}((\tilde\lambda^m) {\cal C}_m^{\;\;\;k}\nabla_k)
(K_{1A}\psi^{A}) ,
\ee
and 
\be
i\sqrt{2}((\tilde\lambda^m)^\dagger {\cal C}_m^{\;\;\;k}\nabla_k)
(K_{2\bar A} \psi^{\bar A}) 
+i\sqrt{2}(\tilde\lambda^m {\cal C}_m^{\;\;\;k}\nabla_k)
(K_{2A}\psi^{A}) .
\ee
Since $J^{(3)}{\cal C} = - {\cal C}J^{(3)}$, the (anti-)holomorphic 
covariant derivatives are now paired with $\tilde\lambda^\dagger$'s 
($\tilde\lambda$'s). As in case of $\bar H_{0,3}$ Yukawa terms,
we can complete the above expression by adding appropriate 
(anti-)holomorphic derivatives of $K_{1,2}$ ($K^*_{1,2}$).  The end
result is,
\be
i\lambda^m({\cal C}J^{(3)})_m^{\;\;\;k}\nabla_k K_{1a}\psi^a
+i\lambda^m{\cal C}_m^{\;\;\;k}\nabla_k K_{2a}\psi^a ,
\ee
which can be rewritten as
\be
i\lambda^m J^{(1)k}_m\nabla_k K_{1a}\psi^a
+i\lambda^m J^{(2)k}_m\nabla_k K_{2a}\psi^a ,
\ee
where we use the fact that $\cal C$ is identified 
with a second complex structure,
$J^{(2)}$, and that $J^{(2)} J^{(3)}$ becomes yet another complex structure,
$J^{(1)}$, completing the triplet of complex structures necessary
for the hyper-K\"ahler geometry.

Adding up all terms, we thus find the following set of fermion bilinears
from the Yukawa terms,
\be
-i\lambda^m \nabla_m K_{0a}\psi^a
+i\sum_{s=1}^3\lambda^m J^{(s)k}_m\nabla_k K_{sa}\psi^a .
\ee
The identity \p{holo} allows us to rewrite this as
\be
-i\lambda^m \nabla_m K_{0a}\psi^a 
-i\sum_{s=1}^3\lambda^m\nabla_m I^{(s)b}_aK_{sb}\psi^a .
\ee
After combining
with the bosonic potential terms derived in the last subsection,
we find the supersymmetric potential terms arising from the matter 
Higgs fields having non-vanishing expectation values is given by
\be\label{lastequation}
-\frac{1}{2}\sum_{i=0}^3|K_i|^2 - i\lambda^m \nabla_m K_{0a}\psi^a
- i\sum_{s=1}^3\lambda^m\nabla_m I^{(s)b}_aK_{sb}\psi^a .
\ee
We will discuss the supersymmetry of the action including these extra potential
terms in the next section.

\section{Supersymmetric Low Energy Dynamics}

For the convenience of the reader, this section summarises 
the general low-energy dynamics of monopoles in $N=2$ Yang-Mills 
theories with hypermultiplets that we have derived. 
We also discuss the quantisation.
Firstly, the action is given by
\bea\label{aaa2}
{\cal L}&=&{1\over 2} \biggl( g_{mn} \dot{z}^m \dot{ z}^n +
ig_{mn} \lambda^m D_t \lambda^n +i\psi^a{\cal D}_t\psi^a  + 
{1\over 2}F_{mn ab}\lambda^m \lambda^n\psi^a\psi^b  \nn
&& - g^{mn} G_m G_n - iD_m G_n  \lambda^m \lambda^n -i T_{ab}\psi^a\psi^b\nn
&& -K_i^a K_{ia} - 2i I_{a}^{(i)\,b} K_{ib;m}\lambda^m\psi^a\biggr),
\eea
where $I_a^{(0)b}=\delta_a^b$, and $i$ runs from 0 to 3.
The action is invariant under $N=4$ supersymmetry transformations given by
\bea
\delta z^m &=& -i\ep\lambda^m +i\ep_s {J^{(s)m}}_n \lambda^n ,\nn
\delta \lambda^m&=&(\dot z^m -G^m)\ep +{J^{(s)m}}_n(\dot z^n -
G^n)\ep_s
-i\ep_s \lambda^k \lambda^n {J^{(s)l}}_k \Gamma^m_{ln} ,\nn
\delta\psi^a&=&-{{A_m}^a}_b\delta z^m\psi^b-\ep (I^{(i)})^a_b K_i^b 
- \ep_s (I^{(i)})^a_b(I^{(s)})^b_c K_i^c ,
\eea
where $\ep,\ep_s$ are constant one component Grassmann odd parameters,
provided that several differential constraints are met: The first is 
the well-known requirements that the moduli space is hyper-K\"ahler
and the curvature $F$ is of (1,1) type with respect to all three
complex structures of the manifold. In addition $G$ must be a tri-holomorphic
Killing vector field, and the two form on the bundle $T$ must satisfy
\bea
T_{ab;m}= F_{mn ab}G^n .
\eea
The section $K$'s on the dual bundle must satisfy a holomorphicity condition
\be
(J^{(s)}\nabla)(I^{(s)}K_i)=\nabla K_i ,\label{holoagain}
\ee
for each $s=1,2,3$, and must also be ``preserved'' under the translation
by $G$
\be
G^m\nabla_m K_{ia}= T_{a}^{\;\;\;b}K_{ib} .\label{GK}
\ee
An additional condition is that
\be\label{abcd}
K_i^aI^{(s)}_{ab} K_j^b=0 .
\ee
When the sections $K$ are non-vanishing we also require
\be\label{abcde}
{(I^{(s)})^c}_b T_{ca}={(I^{(s)})^c}_a T_{cb} .
\ee
We shall show in appendix A, that \p{GK}, \p{abcd} and \p{abcde}
are indeed satisfied.
The action is also invariant under the following symmetry transformation
generated by the tri-holomorphic Killing 
vector field:
\bea
\delta z^m &=&k G^m ,\nn
\delta\lambda^m&=&k{G^m}_{,n}\lambda^n ,\nn
\delta\psi^a&=&k {T^a}_b\psi^b-A^a_{m b}\delta z^m\psi^b , 
\eea
where $k$ is a constant.
This symmetry is responsible for the presence of a central charge in the
superalgebra.

Let us summarize the origin of various terms.
\begin{itemize}
\item
The first line contains the basic ingredient of the monopole dynamics in $N=2$
Yang-Mills theories. The $z^m$ are coordinates on the monopole moduli
space with metric $g_{mn}$. The $\lambda$'s take values 
in the tangent bundle while the $\psi$'s take values in the 
Index bundle of the matter fermions. Generically, this Index bundle 
has a unitary structure group, but for real or pseudo-real matter
representations it is symplectic or orthogonal, respectively.
All interactions are thus
encoded in the geometry of the moduli space and of the Index bundles over it.
These terms suffice if, up to $U(1)_R$ rotation, a single adjoint 
Higgs field has a non-vanishing vacuum expectation value, 
and no other Higgs field does. 

\item
The second line is necessary when the second adjoint Higgs field is turned
on and is not proportional to the first in the Lie algebra space. This
is possible for rank two or higher gauge groups. Extra information is 
contained in the tri-holomorphic Killing vector field $G$, which is picked 
out by the adjoint Higgs expectation values via \p{horse}.
The two-form $T$ is defined via \p{tee} and when the bare
mass for the hypermultiplets is non-vanishing 
it includes a constant piece as discussed in section 3.3.

\item
The third line is necessary when scalar fields in real massless
hypermultiplets get
a vacuum expectation value and still preserves the unbroken $U(1)$
gauge groups. In this case, the Index bundle is symplectic and admits
three covariantly constant complex structures $I^{(s)}$. The sections $K$
must be $G$-invariant in the sense of Eq.~(\ref{GK}), 
and must be holomorphic in the sense of Eq.~(\ref{holoagain}). 
Their normalization is determined by the Higgs
expectation values via \p{fermrel}.

\end{itemize}

An important special case of the above Lagrangian occurs
when one has a single massless adjoint hypermultiplet. The field theory
is then $N=4$ Yang-Mills theory, and the $\psi$'s live in the tangent
bundle. The above Lagrangian should then be the same as the complete monopole
dynamics in $N=4$ Yang-Mills theory, first presented in \cite{blytwo}. 
This can be seen easily by identifying $K$'s as the additional tri-holomorphic 
Killing vector fields\footnote{That the $K$'s must be 
tri-holomorphic Killing vector fields arises from the additional
supersymmetries.} on the moduli space and setting $T_{ab}=G_{a;b}$.

To quantize the effective action we first introduce a frame $e^E_m$ and define
$\lambda^E=\lambda^m e_m^E$ which commute with all bosonic variables.
The remaining canonical commutation relations are then given by
\bea
[z^m,p_n]&=&i\delta^m_n  , \nn
\{\lambda^E,\lambda^F\}&=&\delta^{EF} ,\nn
\{\psi^a,\psi^b\}&=&\delta^{ab} .
\eea
We can realize this algebra on spinors on the moduli space by
letting $\lambda^E=\gamma^F/{\sqrt 2}$, where $\gamma^F$ are gamma matrices.
The states must also provide a representation of the 
Clifford algebra generated
by the $\psi$'s.
The supercovariant momentum operator defined by
\be
\pi_m=p_m-{i\over 4}\omega_{m EF}[\lambda^E,\lambda^F] -{i\over 2}
A_{iab}\psi^a \psi^b,
\ee
where $\omega_{m\, F}^{\, \, E}$ is the spin connection, then becomes
the covariant derivative acting on spinors twisted in
an appropriate way by $A$. Note
that
\bea
{[}\pi_m,\lambda^n{]}&=&i\Gamma^n_{mk}\lambda^k ,\nn
{[}\pi_m,\psi^a{]}&=&i{{A_m}^a}_b\psi^b, \nn
{[}\pi_m,\pi_n{]}&=&
-{1\over 2}R_{mnkl}\lambda^k\lambda^l 
-{1\over 2}F_{mn ab}\psi^a\psi^b .
\eea

The supersymmetry charges take the form
\bea
Q&=&\lambda^m(\pi_m-G_m)-\psi^a \sum_{i=0}^3 (I^{(i)} K_i)_a ,\nn
Q_{s}&=&\lambda^m {J^{(s)n}_{\,\, m}}(\pi_n-G_n) - \psi^a \sum_{i=0}^3 ( 
I^{(i)} I^{(s)}K_{i})_a .
\eea
The algebra of supercharges is given by
\bea\label{algebra}
\{Q,Q\}&=&2({\cal H}-{\cal Z})  , \nn
\{Q_s,Q_t\}&=&2\,\delta_{st}({\cal H}-{\cal Z}) , \nn
\{Q,Q_s\}&=&0 ,
\eea
where the Hamiltonian ${\cal H}$ and the central charge ${\cal Z}$ is given
by
\bea
{\cal H}&=&
{1\over 2\sqrt{g}}\pi_m \sqrt{g }g^{mn}\pi_n
+ {1\over 2}G_m G^m  + {i\over 2}\lambda^m\lambda^n D_m G_n \nn
&& +{i\over 2}
\psi^a \psi^b T_{ab} -{1\over 4}F_{mn ab}\lambda^m\lambda^n\psi^a \psi^b\nn
&& +\frac{1}{2}\, K_i^a K_{ia} +
i I_{a}^{(i)\,b} K_{ib;m}\lambda^m\psi^a ,\nn
{\cal Z}&=& G^m \pi_m -{i\over 2}  \lambda^m\lambda^n(D_m G_n)
+{i\over 2}\psi^a \psi^b T_{ab}.
\label{hamiltonian}
\eea
Note that the operator $i{\cal Z}$ is  the Lie derivative ${\cal L}_G$
acting on spinors twisted by $T$.

Although the algebra of supercharges contains a central
charge $\cal Z$ we see that the states will either preserve all
four supersymmetries of the supersymmetric quantum mechanics if 
${\cal H}={\cal Z}$, or none. 
This is entirely consistent with the fact that the parent
$N=2$ field theory has a complex central charge and hence 
BPS states preserve 1/2 of the eight field theory supercharges,
while generic states preserve none of the supersymmetry (of
course the vacuum preserves all of the supersymmetry). 

\section{Conclusions}

We have presented a detailed derivation of the
effective action governing the low-energy dynamics of 
monopoles and dyons in $N=2$ super-Yang-Mills theory
with hypermultiplets. 
It is valid when both adjoint Higgs fields in
the $N=2$ vector multiplet have non-vanishing expectation
values. We have thus derived the supersymmetric quantum
mechanics presented in \cite{gkpy} and generalised it to
include the effects of the hypermultiplet fermion zero modes.

Our dynamics is also valid for certain cases when it is possible
to have Higgs fields in the hypermultiplets acquire expectation values
while maintaining a non-trivial Coulomb branch. This situation arises
when the matter representation contains a zero weight vector.
Our derivation in section 4 analysed cases when the matter fields are in
real representations. A special case of this is $N=4$ super-Yang-Mills
theory and we have thus derived the supersymmetric quantum mechanics of
\cite{blytwo}. Note that a representation (of a hypermultiplet) does 
not have to be real to have a zero weight vector and it would be interesting 
to know what the dynamics is for this general case. 

It is interesting that the low-energy dynamics of monopoles gives
rise to supersymmetric quantum mechanics that have not been considered
previously. We showed that they can be obtained by a non-trivial
dimensional reduction of $(4,0)$ sigma models in two dimensions.

Finally, it would be interesting to use the effective action to study
the BPS dyon spectrum in more general situations than have been
considered so far. The most promising direction might
be to generalise the approach of \cite{sternyi} using index theorems.

\medskip
\section*{Acknowledgments}
\noindent
JPG is supported in part by an EPSRC Advanced Fellowship
and by PPARC through SPG \#613. KL is supported in part by KOSEF
1998 Interdisciplinary Research Grant
98-07-02-07-01-5.

\section*{Appendix A}
\renewcommand{\theequation}{A.\arabic{equation}}

In this appendix, we derive the conditions \p{holo}, \p{GK} and \p{abcd} 
satisfied by the sections $K_i$ and also establish \p{abcde}.
\vskip 5mm

\leftline{\it  Holomorphicity Condition for $K_i$}
\vskip 5mm

First, we derive the holomorphicity condition (\ref{holo}).
Since $\Psi$ is in a real representation, charge-conjugated zero modes 
can be expressed in terms of original zero modes as in the $\chi$ case,
\be\label{foot}
\gamma_5\Psi^c_{\bar A} = i\tilde C_{\bar A}^{\;\;B} \Psi_B ,
\ee
where $\tilde C$ is an anti-symmetric unitary matrix with $\tilde C^2=-1$. 
Then, using the expansions (\ref{dhep}) and (\ref{dhem}), the 
relations  (\ref{chih}), (\ref{chich}) imply that 
\be  \label{constraint}
\tilde\lambda^{\dagger m} s_m K_i^{\bar A} 
       \tilde C_{\bar A}^{\;\;\;B}\Psi_B + \ldots
  =\tilde\lambda^{\dagger m} C_m^{\;\;\;n} s_n K_i^A\Psi_A + \ldots ,
\ee
where, for simplicity, we omitted the index $i$ in $K_i^A$ which 
plays no role in the appendix. Taking the inner product with 
$\Psi^\dagger_{\bar A}$, we find
\be
\tilde\lambda^{\dagger m} \nabla_m K_i^{\bar A} 
    \tilde C_{\bar A}^{\;\;\;B}
   =\tilde\lambda^{\dagger m} C_m^{\;\;\;n} \nabla_n K_i^A .
\ee
This is a nontrivial condition on $K$. Written in terms of the
real quantities introduced in (\ref{realk}) and (\ref{reallambda}),
it becomes
\be \label{h3}
(1+iJ^{(3)})\nabla I^{(2)}(1-iI^{(3)}) K
    = -(1+iJ^{(3)})J^{(2)}\nabla(1+iI^{(3)}) K ,
\ee
where $I^{(3)}$ is the third complex structure 
of the Index bundle which transforms the real part of $K$ into the 
imaginary part of $K$; $I^{(2)}$ is the second complex structure 
similar to $J^{(2)}={\cal C}$ in the $\chi$ case and has the 
block-diagonal form
\be
I^{(2)} = \pmatrix{ \tilde C & 0 \cr 
                          0  & -\tilde C} ,
\ee
when acting on $\pmatrix{K_{2A-1} \cr K_{2A}}$. All the quantities
are now real so the real and the imaginary parts of (\ref{h3}) should 
hold separately. In fact they reduce to the same condition
\be \label{j3}
\nabla K = J^{(1)} \nabla I^{(1)} K + J^{(2)} \nabla I^{(2)} K
        - J^{(3)} \nabla I^{(3)} K .
\ee

Clearly (\ref{j3}) is consistent with the holomorphicity condition
(\ref{holo}) but is  not exactly the same. More conditions can be obtained by
considering the fermionic zero modes associated with a complex
structure other than $J^{(3)}$. We first generalise \p{fizz}, (\ref{sunday})
by introducing $c$-number spinors $\epsilon_+^{(s)}$, $s=1,2,3$ 
satisfying
\be
\epsilon_+^{(s)\dagger}\epsilon_+^{(s)}=1, \qquad
J^{(s)}_{\mu\nu}=-i\epsilon_+^{(s)\dagger}\Gamma_{\mu\nu}\epsilon_+^{(s)},
\qquad 
J^{(s)}_{\mu\nu} \Gamma_\nu \epsilon_+^{(s)} = i \Gamma_\mu \epsilon_+^{(s)} ,
\ee
and denoting the corresponding zero modes as 
\be
\chi_m^{(s)} = \delta_m W_\mu\Gamma^\mu\epsilon_+^{(s)} .
\ee
The explicit form of $\epsilon_+^{(s)}$ can be found in the following way.
{}From the definition \p{cmn} of $J^{(2)}={\cal C}$, 
\bea
\delta_m W_\mu \Gamma_\mu \epsilon'_+
   &=&C_m^{\;\;\;n} \delta_n W_\mu \Gamma_\mu \epsilon_+ \nn
   &=&-J^{(2)}_{\mu\nu} \delta_m W_\nu \Gamma_\mu \epsilon_+ ,
\eea
from which we find
\be
J^{(2)}_{\mu\nu} \Gamma_\nu \epsilon_+ = \Gamma_\mu \epsilon'_+ .
\ee
(We will continue to omit the superscript label for quantities associated 
with $J^{(3)}$.) Therefore $\epsilon_+^{(2)}$ has the form
\be \label{eps2}
\epsilon_+^{(2)} = \frac{e^{-i\pi/4}}{\sqrt{2}}(\epsilon_+ - i\epsilon'_+) ,
\ee
where the phase is chosen to simplify equations appearing below. 
$\chi_m^{(2)}$ is then given by
\be
\chi_m^{(2)} = \frac{e^{-i\pi/4}}{\sqrt{2}}(1-iJ^{(2)})_m^{\;\;\;n} \chi_n
\ee
With the definition $\epsilon^{'(2)}=C\epsilon_+^{(2)*}$, we can also
expand the complex-conjugated zero modes in terms of $\chi_n^{(2)}$,
\bea
\delta_m W_\mu \Gamma_\mu \epsilon^{'(2)}
&=&\frac{e^{i\pi/4}}{\sqrt{2}}(-i+J^{(2)})_m^{\;\;\;n} \chi_n \nn
&=&(J^{(2)}J^{(3)})_m^{\;\;\;n} \chi_n^{(2)} ,
\eea
where the relation $J_m^{\;\;\;n} \chi_n = i\chi_n$ is used. 
With $J^{(1)}=J^{(2)}J^{(3)}$, the above equation corresponds to the 
counterpart of (\ref{cmn}). A similar analysis can be repeated for $J^{(1)}$, 
but we will omit the details.

Now let us consider the expansion of $\Ds \bar H_i \epsilon_+^{(2)}$,
\be \label{dhep2}
\Ds \bar H_i \epsilon_+^{(2)} 
    \equiv -i\sqrt{2}\gamma_5K_i^{(2)A}\Psi_A .
\ee
{}From the relation (\ref{eps2}), it follows that $K_i^{(2)A}$ is
given by
\be
K_i^{(2)} = \frac{e^{-i\pi/4}}{\sqrt2}(K + i \tilde CK_i^*) .
\ee
In terms of real quantities, this equation becomes
\be \label{k2}
K_i^{(2)} = \frac12( 1+I^{(2)})(1+I^{(3)})K .
\ee
With the expansion (\ref{dhep2}), the condition arising from (\ref{constraint})
now takes the form
\be \label{h2}
(1+iJ^{(2)})\nabla I^{(2)}(1-iI^{(3)}) K_i^{(2)}
    = -(1+iJ^{(2)})J^{(1)}\nabla (1+iI^{(3)})K_i^{(2)} .
\ee
Inserting (\ref{k2}) 
into (\ref{h2}) we obtain a condition for $K$,
\be 
\nabla K_i = J^{(1)} \nabla I^{(1)} K_i - J^{(2)} \nabla I^{(2)} K_i
        + J^{(3)} \nabla I^{(3)} K_i .
\ee
Performing a similar analysis for complex structures $J^{(1)}$
gives 
\be
\nabla K_i = J^{(s)} \nabla I^{(s)} K_i + J^{(t)} \nabla I^{(t)} K_i
         - J^{(u)} \nabla I^{(u)} K_i ,
\ee
where $(s,t,u)$ is a cyclic permutation of $(1,2,3)$. Collectively,
these condition implies the holomorphicity condition \p{holo}.

\vskip 5mm
\leftline{\it Invariance of $K_i$'s under $G$}
\vskip 5mm

To establish \p{GK} consider the following integral:
\be
\frac{1}{\sqrt 2}\int d^3x\Psi^\dagger_{\bar A}\gamma^0 
\, G^m s_m\,\Ds \bar H_i\epsilon_+ .
\ee
After substituting \p{dhem} and directly integrating, this becomes
\be
G^m\nabla_m K_{i\bar A}.
\ee
Alternatively, we can commute $\Ds$ through $G^m s_m$ and integrate by
parts (noting that the surface term vanishes) to get
\be
-\frac{1}{\sqrt 2}\int d^3x D_\mu\Psi^\dagger_{\bar A}
\gamma^0\Gamma_\mu \, G^m s_m\,\bar H_i\epsilon_+
+\frac{1}{\sqrt 2}\int d^3x\Psi^\dagger_{\bar A}\gamma^0 
\, G^m [s_m\, ,\Ds] \bar H_i\epsilon_+ .
\ee
The first term vanishes since $\gamma_5\Psi_A$ is a zero mode while
the second term can
be written using \p{cross} and \p{cat} as 
\bea
\frac{1}{\sqrt 2}\int d^3x\Psi^\dagger_{\bar A}\gamma^0 
\, G^m \delta_m W_\mu\Gamma_\mu 
\bar H_i\epsilon_+ 
= -\frac{1}{\sqrt 2}\int d^3x\Psi^\dagger_{\bar A}\gamma^0 
\,D_\mu\bar a \Gamma_\mu 
\bar H_i\epsilon_+ ,
\eea
which equals
\bea
\frac{1}{\sqrt 2}\int d^3x\Psi^\dagger_{\bar A}\gamma^0 
\,\bar a \Ds \bar H_i\epsilon_+
=K_{i}^B\int d^3x \Psi_{\bar A}^\dagger 
\,\bar a \Psi_B .
\eea
Thus we find
\be
G^m\nabla_m K_{i\bar A} = 
T_{\bar A}^{\;\;\;\bar B}K_{i\bar B} .
\ee
Repeating the exercise for the charge-conjugated version, we find 
\be
G^m\nabla_m K_{ia}= T_{a}^{\;\;\;b}K_{ib} .
\ee

\vskip 5mm
\leftline{\it Vanishing of $\langle K_i|I^{(s)}|K_j\rangle$}
\vskip 5mm

Consider the simplest case of $I^{(3)}$. From the definition of the $K$'s,
this inner product is equal to the integral
\be
-\frac{i}{2}\int d^3x\;\left( \epsilon_+^\dagger \Ds \bar H_i \Ds \bar H_j\epsilon_+
-(\epsilon_+')^\dagger \Ds \bar H_i \Ds \bar H_j\epsilon_+'\right) .
\ee
Since $\Ds^2 H\epsilon_+=\Ds^2 H\epsilon'_+=0$ this is a boundary 
integral given by
\be
-\frac{i}{2}\oint \;d\hat n_\mu \bar H_i D_\nu \bar H_j
\left( \epsilon_+^\dagger \Gamma_\mu\Gamma_\nu \epsilon_+ -
(\epsilon_+')^\dagger \Gamma_\mu\Gamma_\nu \epsilon_+'\right) .
\ee
Since $\epsilon_+$ and $\epsilon_+'$ are normalized to unity, the symmetric
part of $\Gamma_\mu\Gamma_\nu$ in each term cancel. The anti-symmetric part
is proportional to $J^{(3)}_{\mu\nu}$ which is a complex structure on $R^4$;
\be
\oint \;d\hat n_\mu \bar H_i D_\nu \bar H_j J^{(3)}_{\mu\nu} .
\ee
This integrand consists of angular covariant derivatives of $\bar H_j$ contracted
with $\bar H_i$ as well as a term involving the action 
of the adjoint field $b$ on $\bar H_j$ again contracted with $\bar H_i$. 
Lets work in the unitary gauge where the unbroken gauge
$U(1)$ generators are taken to be diagonal.
In the asymptotic region the only surviving 
terms are then {\it ordinary} angular derivatives on $\bar H_j$, since all the
other terms are exponentially small and do not contribute to the surface 
integral.

Since $\bar H_j$ must solve the {\it ordinary} 3-dimensional 
Laplace equation at large $r$ its asymptotic form is given by
\be
\bar H_j=\langle \bar H_j\rangle+\sum_{lm} c_{lm}
\frac{Y_{lm}}{r^{l+1}}+\cdots ,
\ee
where $c_{lm}$ are constant vectors, $Y_{lm}$ are the 
3-dimensional spherical harmonics and
the ellipsis denotes terms that are exponentially small in large 
$r$. Since the coefficient of the leading $1/r$ piece, $Y_{00}$, 
is a constant, the boundary integral vanishes on the asymptotic two-sphere.
Similar consideration starting with
different $\epsilon_+$ as in the derivation of the holomorphicity 
condition above leads us to 
\be
\langle K_i|I^{(s)}|K_j\rangle =0 .
\ee
for $s=1,2,3$.

\vskip 5mm
\leftline{\it Establishing ${(I^{(s)})^c}_b T_{ca}={(I^{(s)})^c}_a T_{cb}$}
\vskip 5mm
For $I^{(3)}$ this condition is equivalent to the statement that
$T_{AB}$=$T_{\bar A\bar B}=0$ which is true by definition. For $I^{(2)}$
consider the term with $a,b$ both being holomorphic
indices $A,B$. We then have
\bea
{I^{(2)\bar C}}_B T_{\bar C A}&=&\int d^3 x{I^{(2)\bar C}}_B \Psi^\dagger_{\bar C} 
\bar a\Psi_A\nn
&=&\int d^3 x \psi^T_B  C\bar a\psi_A
\eea
where we have used \p{foot}. This is symmetric in $A,B$ since both $C$ and
the group generators are anti-symmetric. Other components and $I^{(1)}$
can be dealt with similarly.

\section*{Appendix B}  

\renewcommand{\theequation}{B.\arabic{equation}}
\setcounter{equation}{0}
The supersymmetric quantum mechanics (\ref{aaa2}), which
generalises that presented in \cite{gkpy}, is as far as we 
know new. We show here that it can be obtained from a
non-trivial, ``Scherk-Schwarz'', dimensional reduction of 
a two-dimensional sigma model with (4,0) supersymmetry. 

Let $(\sigma^0,\sigma^1,\theta^+)$ be coordinates of 
two-dimensional (1,0) superspace and consider the following  action
\be\label{twodee}
{\cal S}={1\over 2}\int d^2\sigma d\theta^+\left[ iD_+
z^m\partial_=z^n g_{mn}+
\psi^a_-\nabla_+ \psi^b_- h_{ab}\right] .
\ee
Here
$\sigma^{\neq}=(\sigma^0+\sigma^1)/2$, $\sigma^==(\sigma^0-\sigma^1)/2$ and
$D_+=\partial_{\theta^+}-i\theta^+\partial_{\neq}$. The scalar superfield
$z^m$ is a map from (1,0) superspace to a target ${\cal M}$ 
and the Grassmann odd 
superfield $\psi^a_-$ takes values in a vector bundle over $\cal M$. 
$h_{ab}$ is a fiber metric satisfying $\nabla_ih_{ab}=0$
and $\nabla_+\psi^a_-=
D_+\psi^a_-+{{A_m}^a}_b D_+z^m\psi^b_-$, where $A$ is a connection
on the vector bundle. 

The component form of the action can be obtained by first
expanding the superfields via
\bea
z^m&=&z^m+i\theta^+\lambda^m_+\nn
\psi^a_-&=&\psi^a_-+\theta^+ f^a .
\eea
After eliminating the auxiliary
fields $f$ via their equations of motion we obtain
\be
{\cal S}={1\over 2}\int d^2l [\partial_{\neq}z^m\partial_=
z^n g_{mn}
+i \lambda^m_+\nabla_=\lambda^n_+g_{mn} + 
i \psi^a_-\nabla_{\neq}\psi^b_-h_{ab}
+{1\over 2}F_{mn ab}\lambda^m_+\lambda^n_+\psi^a_-\psi^b_-] ,
\ee
where $F$ is the curvature of the connection $A$ and $\nabla_=$ and 
$\nabla_{\neq}$ are the covariantization of $\partial_=$ and 
$\partial_{\neq}$, respectively, with the pull back of the
Christoffel symbols.

Let us suppose that the target manifold 
is hyper-K\"ahler and that the connection 
is tri-holomorphic so that
the sigma model admits an extended (4,0)
supersymmetry.
Suppose in addition that 
the action is invariant under the symmetry transformations
generated by a tri-holomorphic Killing vector field $G^m$:
\bea\label{symone}
\delta z^m&=&kG^m ,\nn
\delta \psi^a_-&=&k{T^a}_b\psi^a_- -{{A_m}^a}_b\delta z^m\psi^b_- ,
\eea
where $k$ is a constant and the tensor $T_{ab}=-T_{ba}$ must satisfy
\be\label{symcond}
G^k F_{k m ab}=-T_{ab;m} ,
\ee
which determines $T$ up to covariantly constant terms.

Ordinary dimensional reduction to a supersymmetric quantum mechanics 
is implemented by assuming that 
all of the fields are independent of the coordinate
$\sigma^1$. Scherk-Schwarz reduction is achieved by demanding the
weaker condition that the Lagrangian is independent. Using the
invariance under the symmetry transformations \p{symone} this can be
achieved by letting the $\sigma^1$
dependence of the fields be given by
\bea
\partial_1z^m&=&-G^m ,\nn
\partial_1\lambda^m_+&=&-{G^m}_{,n}\lambda^n_+ ,\nn
\partial_1\psi^a_-&=&-{T^a}_b\psi^b_-+{{A_m}^a}_bG^m\psi^b_- .
\eea
After integrating over $\sigma^1$ one then obtains the following action
\bea\label{pig}
S&=&{1\over 2}\int dt \big[\dot z^m\dot z^n g_{mn}
-G^m G^n g_{mn} +i\lambda^m D_t\lambda^n_+ g_{mn} 
+i\lambda^m_+\lambda^n_+ G_{m;n} \nn
&&+i\psi^a_-{\cal D}_t\psi^b_- h_{ab}- i \psi^a_- \psi^b_- T_{ab}
+{1\over 2}F_{mn ab}\lambda^m_+\lambda^n_+\psi^a_-\psi^b_-\big] .
\eea
Identifying $\lambda_+$'s with $\lambda$'s,
and $\psi_-$'s with $\psi$'s, we recover precisely the 
effective action \p{action} that describes the
dynamics of monopoles with fermionic contributions from the
hypermultiplets.

To obtain the supersymmetric quantum mechanics when the hypermultiplets
have non-zero expectation values, we generalise the above
construction\footnote{Another generalization,
is to reduce a model with torsion $H=db$. In the case that the Lie-derivative
with respect to $G$ of the two-form $b$ vanishes the Scherk-Schwarz
reduction proceeds in a straightforward manner. We do not present any details
here as there is no obvious application to monopole
dynamics.} by performing Scherk-Schwarz reduction on a 
(4,0) model with potential \cite{hpt}.
To do this we add a (1,0) supersymmetric term, 
\be\label{newbit}
\Delta {\cal S} =\frac{1}{2}\int d^2\sigma d\theta^+\;2  v_a \psi^a_- ,
\ee
to \p{twodee} where $v_a$ is a section of the dual of the Index bundle. 
The combined quantum mechanics 
action is invariant under the symmetry transformations
\p{symone}
provided that in addition to \p{symcond} the section satisfies 
\be
v^aT_{ab}+v_{b;k}G^k=0 .
\ee
Since the action \p{newbit} does not contain any derivatives, the
Scherk-Schwarz reduction is equivalent to ordinary dimensional reduction. 
After eliminating the auxiliary fields it leads to 
\be
\Delta {S} =\frac{1}{2}\int dt [ - v_a v_b h^{ab}  +2
i v_{a;m}\lambda^m\psi^a ] .
 \ee
The combined action is automatically invariant under 
an $N=1$ supersymmetry.
The extended $N=4$ supersymmetry of \p{pig} will
extend to that of the combined action under suitable 
conditions on the section $v^a$. The supersymmetry transformations are
\bea
\delta z^m &=& -i\ep\lambda^m +i\ep_s {J^{(s)m}}_n \lambda^n ,\nn
\delta \lambda^m&=&(\dot z^m -G^m)\ep +{J^{(s)m}}_n(\dot z^n -
G^n)\ep_s
-i\ep_s \lambda^k \lambda^n {J^{(s)l}}_k \Gamma^m_{ln} ,\nn
\delta\psi^a&=&-{{A_m}^a}_b\delta z^m\psi^b + \ep v^a + \ep_s t_{(s)}^a ,
\eea
provided that the sections $t^a_{(s)}, s=1,2,3$ can be found 
satisfying
\bea
(v_a t^a_{(s)})_{;m}&=&0 ,\nn
J^{(s)n}_m \nabla_n v^a&=&-\nabla_m t^a_{(s)} ,\nn
G^n t^a_{(s);n}&=& T_{ab} t^b_{(s)} ,
\eea
where  $J^{(s)}$ are the three complex structures on the target manifold. 
Note that these conditions imply that the norm of $v$ and those of the
$t$'s differ only by a constant.

Consider now the particular case that the bundle associated with
the fermionic variables $\psi^a$ has the structure group 
$Sp(n)$. In this case there exists four covariantly constant rank-two
tensors; the identity $I^{(0)}$, and the three complex structures
$I^{(s)}$. 
Let us write the section in terms of four vector fields $K_i$ via
\be
v^a=-\sum_{i=0}^3 {(I^{(i)})^a}_b K_i^b .
\ee
A solution for the three related sections $t^a_{(s)}$ is
\be
t^a_{(s)}=-\sum_{i=0}^3 {(I^{(i)})^a}_b{(I^{(s)})^b}_c K_i^c ,
\ee
providing that the $K_i$'s satisfy
\bea
J_m^{(s)k}\nabla_k(I_a^{(s)b}K_{ib}) &=& \nabla_m K_{ia} 
\qquad (\hbox{no sum on $s$}) ,\nn
K_i^aI^{(s)}_{ab} K_j^b&=&\hbox{constant} ,\nn
G^n K_{ia;n}&=&  T_{ab}   K_i^b ,\nn
{(I^{(s)})^c}_b T_{ca}&=&{(I^{(s)})^c}_a T_{cb} .
\eea
Note that the first
equation implies that (anti-)holomorphic covariant derivative of 
(anti-)holomorphic part of $K_i$ vanishes. 
If we make further assumption the constants $K_i^aI^{(s)}_{ab} K_j^b$
actually vanishes, the supersymmetric potential terms are given by
\be
-\frac{1}{2}\sum_{i=0}^3 \left(
K_i^a K_{ia} +2i I_{a}^{(i)\,b} K_{ib;m}\lambda^m\psi^a\right) ,
\ee
which are precisely those arising from hypermultiplet 
vacuum expectation values that we established in section 4.



\medskip


\begin{thebibliography}{99}

\bibitem{gaunt}
J.P.~Gauntlett,
Nucl.\ Phys.\ {\bf B411} (1994) 443, hep-th/9305068.

\bibitem{blum}
 J. Blum, Phys. Lett. {\bf B333} (1994) 92, hep-th/9401133.

\bibitem{sen}
A. Sen, Phys.Lett.{\bf  B329} (1994) 217, hep-th/9402032.

\bibitem{sethi}
S. Sethi, M. Stern and E. Zaslow, Nucl. Phys. {\bf B457} (1995) 484,
hep-th/9508117.

\bibitem{cederwall}
M.~Cederwall, G.~Ferretti, B.~E.~Nilsson and P.~Salomonson,
Mod.\ Phys.\ Lett.\  {\bf A11} (1996) 367,
hep-th/9508124.

\bibitem{gauntharv}
J. P. Gauntlett and J. A. Harvey,
Nucl. Phys. {\bf B463} (1996) 287, hep-th/9508156.

\bibitem{gauntlowe}
J.P. Gauntlett and D.A. Lowe, Nucl. Phys. {\bf B472} (1996) 194,
hep-th/9601085

\bibitem{lwy}
K. Lee, E.J. Weinberg, and P. Yi, Phys. Lett. {\bf B376} (1996) 97,
hep-th/9601097; Phys. Rev. {\bf D54} (1996) 1633, hep-th/9602167.

\bibitem{gibbons}
G. W. Gibbons, Phys. Lett. {\bf B382} (1996) 53, hep-th/9603176.

\bibitem{bergman} O. Bergman, 
Nucl. Phys. {\bf B525} (1998) 104, hep-th/9712211.

\bibitem{hashimoto}
K. Hashimoto, H. Hata, and N. Sasakura,
Phys. Lett. {\bf B431} (1998) 303, hep-th/9803127.

\bibitem{bergmankol}
O. Bergman and B. Kol,
Nucl. Phys. {\bf B536} (1998) 149, hep-th/9804160.

\bibitem{kawano}
T. Kawano and K. Okuyama, Phys.Lett. B432 (1998) 338,
hep-th/9804139.

\bibitem{hhs}
K. Hashimoto, H. Hata, and N. Sasakura,
Nucl.Phys. {\bf B535} (1998) 83, hep-th/9804164.

\bibitem{leeyi}
K.~Lee and P.~Yi,
Phys.\ Rev.\ {\bf D58} (1998) 066005,
hep-th/9804174.

\bibitem{bak}
D. Bak, K.  Hashimoto, B. Lee, H. Min, and N. Sasakura, 
Phys.Rev. {\bf D60} (1999) 046005, hep-th/9901107.

\bibitem{tong}
D.~Tong,
Phys.\ Lett.\ {\bf B460} (1999) 295,
hep-th/9902005.

\bibitem{blly}
D.~Bak, C.~Lee, K.~Lee and P.~Yi,
Phys.Rev. {\bf D61} (2000) 025001, hep-th/9906119; 

\bibitem{blyone}
D.~Bak, K.~Lee and P.~Yi, 
Phys.Rev. {\bf D61} (2000) 045003, hep-th/9907090.

\bibitem{baklee}
D. Bak and K. Lee, 
Phys.Lett. {\bf B468} (1999) 76, hep-th/9909035.

\bibitem{blytwo}
D. Bak, K. Lee, P. Yi, {\it Complete Supersymmetric Quantum Mechanics of
Magnetic Monopoles in $N=4$ SYM Theory}, hep-th/9912083.

\bibitem{gkpy}
J.~P.~Gauntlett, N.~Kim, J.~Park and P.~Yi,
Phys.Rev. {\bf D61} (2000) 125012, hep-th/9912082.

\bibitem{lyexact}
K. Lee and P.Yi,
Nucl.Phys. {\bf B520} (1998) 157, hep-th/9706023

\bibitem{fh}
C.~Fraser and T.J.~Hollowod,
Phys.\ Lett.\ {\bf B402} (1997) 106,
hep-th/9704011.

\bibitem{wittenolive}
E.~Witten and D.~Olive,
Phys.\ Lett.\ {\bf 78B} (1978) 97.

\bibitem{harvstrom}
J.~A.~Harvey and A.~Strominger,
Commun.\ Math.\ Phys.\  {\bf 151} (1993) 221, hep-th/9108020.

\bibitem{callias}
C.~Callias,
Commun.\ Math.\ Phys.\  {\bf 62} (1978) 213.

\bibitem{mantonschroers}
N.~S.~Manton and B.~J.~Schroers,
Annals Phys.\  {\bf 225} (1993) 290.

\bibitem{henning}
M. Henningson, 
Nucl.Phys. {\bf B461} (1996) 101,
hep-th/9510138.

\bibitem{hpt}
C.~M.~Hull, G.~Papadopoulos and P.~K.~Townsend,
Phys.\ Lett.\  {\bf B316} (1993) 291, hep-th/9307013.


\bibitem{sternyi}
M.~Stern and P.~Yi,
{\it Counting Yang-Mills dyons with index theorems},
hep-th/0005275.




\end{thebibliography}
\end{document}